\documentclass[aps,pra,preprint,superscriptaddress,showpacs,showkeys]{revtex4}
\usepackage{amssymb,bm}
\usepackage{graphicx}
\usepackage{amsmath}
\usepackage{epstopdf}
\allowdisplaybreaks

\begin{document}

\title{High-energy  $e^+e^-$ photoproduction  in the  field of a heavy atom accompanied by  bremsstrahlung}

\author{P.A. Krachkov}\email{peter_phys@mail.ru}
\affiliation{Budker Institute of Nuclear Physics, 630090 Novosibirsk, Russia}
\affiliation{Novosibirsk State University, 630090 Novosibirsk, Russia}
\author{R.N. Lee}
\email{R.N.Lee@inp.nsk.su}\affiliation{Budker Institute of Nuclear Physics, 630090 Novosibirsk, Russia}
\author{A. I. Milstein}\email{A.I.Milstein@inp.nsk.su}
\affiliation{Budker Institute of Nuclear Physics, 630090 Novosibirsk, Russia}

\date{\today}

\begin{abstract}
Helicity amplitudes and differential cross section of  high-energy  $e^+e^-$ photoproduction  accompanied by  bremsstrahlung in the electric field of a heavy atom are derived. The results are exact in the nuclear charge number and obtained in the leading quasiclassical approximation. They correspond to the leading high-energy small-angle asymptotics of the amplitude. It is shown that, in general, the Coulomb corrections essentially modify the differential cross section as compared to the Born result. When the initial photon is circularly polarized the Coulomb corrections lead to the asymmetry in the distribution over the azimuth angles $\varphi_i$ of produced particles with respect to the replacement $\varphi_i\to -\varphi_i$.
\end{abstract}



\maketitle



\section{Introduction}

QED processes at high energy in the field of a heavy nucleus or atom are the classical examples of the processes in a strong field. They show up in many experimental setups, including those designed for completely different purposes, not connected with observation of these processes. Therefore, their investigation clearly has a practical value. From the theoretical point of view, these processes are interesting because they provide an important insight into the structure of the higher-order effects of the perturbation theory.

General approach to the strong-field calculations is the use of the Furry representation. In this approach the wave functions and propagators of particles are replaced by the exact solutions and Green functions of the wave equations in the external field. However, even for the pure Coulomb field these objects are very complicated and their use for the practical calculations is limited. Fortunately, at high energies of initial particles the final particle momenta usually have small angles with respect to the incident direction. This is where the quasiclassical approximation comes into play. In this approximation, the wave functions and propagators acquire remarkably simple forms which allow for the effective use in specific calculations. The quasiclassical Green's function of the Dirac equation in the external field have been derived for a number of field configurations, see Ref. \cite{MS1983} for the case of a pure Coulomb field, Ref. \cite{LM95A} for an arbitrary spherically symmetric field, Ref. \cite{LMS00} for a localized field which generally possesses no spherical symmetry, and Ref. \cite{DM2012} for combined strong laser and atomic fields. Even more surprising is the fact that within this approximation it appears to be possible to derive the results not only in the leading order, but also a first quasiclassical correction to them.

Basic processes in the field of heavy atom are the electron-positron pair photoproduction (PP) and electron bremsstrahlung (BS). They both have a long history of investigation, for the former process see reviews in Refs. \cite{HGO1980,Hubbell2000}. For the total cross section of electron-positron pair photoproduction there is also a formal expression \cite{Overbo1968}, exact in the parameter $\eta=Z\alpha$ and the photon energy $\omega$  (here $Z$ is the atomic charge number, $\alpha$ is the fine-structure constant, $\hbar=c=1$). It has the form of multiple slowly converging sums containing the hypergeometric function of two arguments $F_2$. Due to these complications, the computation based on this expression rapidly becomes intractable with the growth of $\omega$, and the numerical results have been obtained so far only for $\omega<12.5\,$ MeV \cite{SudSharma2006}. At high energy the quasiclassical approximation is applicable and the leading  quasiclassical term, for both pair production and bremsstrahlung, has been obtained in  \cite{BM1954,DBM1954,OlsenMW1957,O1955,OM1959}. The first quasiclassical corrections to the spectra of both processes as well as to the total cross section of pair production have been obtained in Refs. \cite{LMS2004,LMSS2005}. It is remarkable that the quasiclassical correction to the total cross section of pair production can not be obtain by simply integrating the quasiclassical correction to the spectrum. This is because of the contribution of the tip regions of the spectrum, where only one particle can be considered quasiclassically. A detailed investigation of this region was made in Ref. \cite{DM10}. The corresponding angular distribution was derived in Ref.~\cite{DM12}. Recently, the  first quasiclassical correction to the fully differential cross section was obtained in Ref. \cite{LMS2012} for $e^+e^-$  pair photoproduction and in Ref. \cite{DLMR2014} for $\mu^+\mu^-$ pair  photoproduction. As a result, charge asymmetry in these processes was predicted.

In the present paper we apply the quasiclassical approach to the investigation of  $e^+e^-$ photoproduction  in the  field of a heavy atom accompanied by  bremsstrahlung. The cross section of this process is a significant part of the radiative corrections to  $e^+e^-$ photoproduction as well as noticeable background to such processes as Delbr\"uck scattering and others. This process should be taken into account at the consideration of the electromagnetic showers in the matter. In spite of its importance, there are only few theoretical results related to this process \cite{Huld1967,Corbo1978}. In those papers the Born approximation was used, while there are no theoretical results exact in the parameter $\eta$. The goal of the present paper is twofold. First, we would like to fill the gap in the theoretical description of the process, and, in particular, determine the magnitude of the Coulomb corrections for various kinematic regions. We show that, apart from the region of very small momentum transfer, the Coulomb corrections for heavy atoms drastically change the result compared to the Born approximation. Second, we would like to demonstrate how smoothly the quasiclassical approach works for this complicated case. We consider in detail the case of a pure Coulomb field and then present the modification due to screening by atomic electrons.


\section{General discussion}

The main contribution to the cross section of the process $\gamma_1Z\to e^+e^-\gamma_2\,Z$ is given by the region of small angles between the momenta of the incoming and outgoing particles. In this region
\begin{equation}\label{eq:cs}
d\sigma=\alpha^2|M|^{2}\frac{d\bm{p}_\perp\,d\bm{q}_\perp\,d\bm{k}_{2\perp}d\varepsilon_pd\varepsilon_q}{(2\pi)^6\omega_1\omega_2}\,,
\end{equation}
where $\bm k_1$, $\bm k_2$, $\bm p$, $\bm q$  are the momenta of initial photon, final photon, electron and positron, respectively, $\varepsilon_{ p}=\sqrt{ \bm{p}^2+m^2}$, $\varepsilon_{ q}=\sqrt{ \bm q^2+m^2}$, $\omega_2=\omega_1-\varepsilon_{ p}-\varepsilon_{ q}$. We fix the coordinate system so that $\bm\nu=\bm k_1/\omega_1$  is directed along $z$-axis, $\bm k_2$ lies in the $xz$ plane with $k_{2x}>0$, the notation $\bm X_\perp=\bm X -(\bm X\cdot\bm \nu)\bm \nu$ for any vector $\bm X$ is used.

The matrix element $M$ has the form
\begin{multline}\label{MG}
M=M^{(1)}+M^{(2)}=-\int d\bm r_1\,d\bm r_2 \,\bar{u}_{\bm p }^{(-)}(\bm r_1 )\Bigg\{(\bm\gamma\cdot
\bm e_{2}^*) e^{-i\bm k_2\cdot\bm r_1}G(\bm r_1,\,\bm r_2|\varepsilon_{ p}+\omega_2) e^{i\bm k_1\cdot\bm r_2}\,(\bm\gamma\cdot
\bm e_{1})\\
+(\bm\gamma\cdot\bm e_{1}) e^{i\bm k_1\cdot\bm r_1}G(\bm r_1,\,\bm r_2|-\varepsilon_{ q}-\omega_2) e^{-i\bm k_2\cdot\bm r_2}\,(\bm\gamma\cdot
\bm e_{2}^*)\Bigg\}\,v _{\bm q}^{(+)}(\bm r_2 )\,.
\end{multline}
\begin{figure}
\centering
\includegraphics[width=0.8\linewidth]{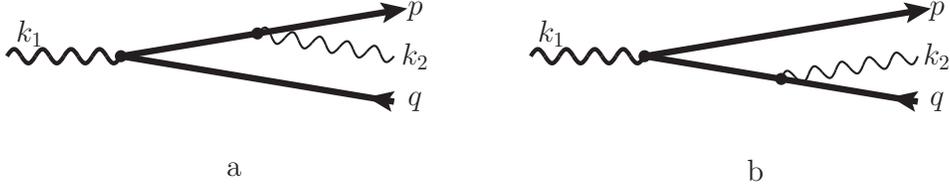}
\caption{Diagrams of the process $\gamma_1Z\to e^+e^-\gamma_2Z$. Thick solid lines denote exact propagators in the nuclear field.}
\label{fig:diagrams}
\end{figure}

Here $ u_{\bm p}^{(-)}(\bm r )$ and $v_{\bm q}^{(+)}(\bm r )$ are the positive- and negative-energy solution of the Dirac equation in the external field, $\bm e_{1}$ and $\bm e_{2}$ are the  polarization vectors of the initial and final photons, respectively,  $\gamma^\mu$ are the Dirac matrices, $G(\bm r_1,\,\bm r_2|\varepsilon)$ is the Green's function of the Dirac equation in the external field. The superscripts $(-)$ and $(+)$ remind that the asymptotic form of $ u_{\bm p}^{(-)}(\bm r )$ and $ v_{\bm q}^{(+)}(\bm r )$ at large $\bm r$ contains, in addition to the plane wave, the spherical convergent and divergent waves, respectively. The first term in Eq. (\ref{MG}), $M^{(1)}$, corresponds to radiation from the electron line and the second term, $M^{(2)}$, corresponds to that from the positron line, see Fig.\ref{fig:diagrams}, a and b, respectively. It is convenient to write Eq. (\ref{MG}) in terms of the Green's function $D(\bm r_1,\,\bm r_2|\varepsilon)$ of the  ``squared'' Dirac equation,
\begin{equation}\label{FGD}
G(\bm r_1,\,\bm r_2|\varepsilon)=(\hat{\cal P}+m)D(\bm r_1,\,\bm r_2|\varepsilon),\quad D(\bm r_1,\,\bm r_2|\varepsilon)=\langle \bm r_1|\frac{1}{\hat{\cal P}^2-m^2+i0}| \bm r_2\rangle,
\end{equation}
$\hat{\cal P}=\gamma^\mu{\cal P}_\mu$, ${\cal P}_\mu=(\varepsilon-V(r),i\bm\nabla)$, and $V(r)$ is the atomic potential. Substituting Eq. \eqref{FGD} in Eq. \eqref{MG}, performing integration by parts and using the Dirac equation, we obtain
\begin{multline}\label{MD}
M=-\,\int d\bm r_1\,d\bm r_2 \,\bar u_{\bm p }^{(-)}(\bm r_1 )\Bigg\{ e^{-i\bm k_2\cdot\bm r_1}[(\bm\gamma\cdot\bm e_{2}^*)\hat{k}_2+2(\bm e_{2}^*\cdot\bm p_1)]
D(\bm r_1,\,\bm r_2|\varepsilon_{ p}+\omega_2)
e^{i\bm k_1\cdot\bm r_2}\,(\bm\gamma\cdot\bm e_{1})\\
+(\bm\gamma\cdot\bm e_{1}) e^{i\bm k_1\cdot\bm r_1}D(\bm r_1,\,\bm r_2|-\varepsilon_{ q}-\omega_2)e^{-i\bm k_2\cdot\bm r_2}\,[(\bm\gamma\cdot
\bm e_{2}^*)\hat{k}_2+2(\bm e_{2}^*\cdot\bm p_2)]\Bigg\}\,v _{\bm q}^{(+)}(\bm r_2 )\,.
\end{multline}
Here $\bm p_1=-i\partial/\partial\bm r_1$, and  $\bm p_2=-i\partial/\partial\bm r_2$.   We first calculate the  term $M^{(1)}$ and then find  $M^{(2)}$   by means of the C-parity transformation.

As it was shown in Ref. \cite{DLMR2014}, the wave functions and the Green's function can be represented in the form
\begin{align}\label{wfD1}
\bar u_{\bm p }^{(-)}(\bm r_1 )&=\bar u_{\bm p }[f_0(\bm r_1,\bm p)-\bm\alpha\cdot\bm f_1(\bm r_1,\bm p)-\bm\Sigma\cdot\bm f_2(\bm r_1,\bm p)]\,,\\
v _{\bm q}^{(+)}(\bm r_2 )&=[g_0(\bm r_2,\bm q)+\bm\alpha\cdot\bm g_1(\bm r_2,\bm q)+\bm\Sigma\cdot\bm g_2(\bm r_1,\bm p)]v _{\bm q}\,,\nonumber\\
D(\bm r_1,\,\bm r_2|\varepsilon)&=[d_0(\bm r_1,\bm r_2)+\bm\alpha\cdot\bm d_1(\bm r_1,\bm r_2)+\bm\Sigma\cdot\bm d_2(\bm r_1,\bm p)]\,,
\end{align}
where
\begin{eqnarray}\label{uv}
 u_{\bm p}=\sqrt{\frac{\varepsilon_p+m}{2\varepsilon_p}}
\begin{pmatrix}
\phi\\
\dfrac{{\bm \sigma}\cdot {\bm
p}}{\varepsilon_p+m}\phi\end{pmatrix}\, ,\quad
v_{\bm q}=\sqrt{\frac{\varepsilon_q+m}{2\varepsilon_q}}
\begin{pmatrix}
\dfrac{{\bm \sigma}\cdot {\bm
q}}{\varepsilon_q+m}\chi\\
\chi\end{pmatrix}\, ,
\end{eqnarray}
and $f_0$, $\bm f_{1,2}$, $g_0$, $\bm g_{1,2}$,  $d_0$, $\bm d_{1,2}$ are some functions, $\phi$ and $\chi$ are spinors. In the quasiclassical approximation the relative magnitude of these functions is different, so that
\begin{equation}
f_0\sim l_c f_1\sim l^2_c f_2\,,\quad
g_0\sim l_c g_1\sim l^2_c g_2\,,\quad
d_0\sim l_c d_1\sim l^2_c d_2\,,
\end{equation}
where $l_c\sim \omega/\Delta\gg 1$ is the characteristic value of the angular momentum in the process, $\bm\Delta=\bm p+\bm q+\bm k_2-\bm k_1$ is the momentum transfer. Nevertheless, it appears that, due to cancellations in the matrix element $M$, it is necessary to keep not only the leading terms $f_0\,,\ g_0\,,\ d_0$, but also the subleading terms $\bm f_1\,,\ \bm g_1\,,\ \bm d_1$, while the terms $\bm f_2\,,\ \bm g_2\,,\ \bm d_2$ can be safely omitted in the leading approximation. Thus, we can write the term $M^{(1)}$ as follows
\begin{multline}\label{MD1}
M^{(1)}=-\int d\bm r_1\,d\bm r_2\, \mathrm{Sp}\{(f_0-\bm\alpha\cdot\bm f_1)[(\bm\gamma\cdot\bm e_{2}^*)\hat{k}_2+2(\bm e_{2}^*\cdot\bm p_1)]
 e^{-i\bm k_2\cdot\bm r_1}\\
\times (d_0+\bm\alpha\cdot\bm d_1)e^{i\bm k_1\cdot\bm r_2}(\bm\gamma\cdot\bm e_{1})(g_0+\bm\alpha\cdot\bm g_1)v _{\bm q}\bar u_{\bm p }\}\,.
\end{multline}
In what follows we calculate the matrix element for definite helicities of the particles. Let $\lambda_1\,,\ \lambda_2\,,\ \mu_p$ and $\mu_q$ be the signs of the helicities of initial photon, final photon, electron, and positron, respectively.
Denoting helicities by the subscripts, we have
\begin{gather}
v _{\bm q\mu_q}\bar u_{\bm p\mu_p}
=\frac{1}{8}(a_{\mu_p\mu_q}+\bm\Sigma\cdot \bm b_{\mu_p\mu_q})[\gamma^0(Q+P)+\gamma^0\gamma^5(1+PQ)-(P-Q)-\gamma^5(1-PQ)],\nonumber\\
\label{calf}
P=\frac{\mu_pp}{\varepsilon_p+m}\,,\quad Q=-\frac{\mu_qq}{\varepsilon_q+m}\,,
\end{gather}
where $a_{\mu_p\mu_q}$ and $\bm b_{\mu_p\mu_q}$ are defined from
\begin{equation}
\chi_{\mu_q}\phi_{\mu_p}^\dagger=\frac{1}{2}(a_{\mu_p\mu_q}+\bm\sigma\cdot \bm b_{\mu_p\mu_q})\,.
\end{equation}
Note that only the terms with  $(P+Q)$ and $(1+PQ)$ in Eq. \eqref{calf} contribute to the matrix element (\ref{MD1}) because it contains the odd  number of the gamma-matrices.

Let us fix the overall phase of the helicity amplitudes by choosing
\begin{align}\label{spinors}
\phi_{\mu_p}&=\frac{1+\mu_p\bm \sigma\cdot\bm n_p}{4\cos(\theta_{p}/2)}
\begin{pmatrix}1+\mu_p\\1-\mu_p\end{pmatrix}
\approx \frac14\left(1+\frac{\theta_{p}^2}{8}\right)\left(1+\mu_p\bm \sigma\cdot\bm n_p\right)\begin{pmatrix}1+\mu_p\\1-\mu_p\end{pmatrix}\,,
\nonumber\\
\chi_{\mu_q}&=-\frac{1-\mu_q\bm \sigma\cdot\bm n_q}{4\cos(\theta_{q}/2)}
\begin{pmatrix}
\mu_q-1\\
\mu_q+1\end{pmatrix}
\approx-\frac{1}{4}\left(1+\frac{\theta_{q}^2}{8}\right)(1-\mu_q\bm \sigma\cdot\bm n_q)
\begin{pmatrix}
\mu_q-1\\
\mu_q+1
\end{pmatrix}\,,\nonumber\\
\bm e_{1\lambda_1}&=\bm e_{\lambda_1}=\frac{1}{\sqrt{2}}(\bm e_x+i\lambda_1\bm e_y)\,,\quad
\bm e_{2\lambda_2}=\frac{1}{\sqrt{2}}(\bm e_x^\prime+i\lambda_2\bm e_y)\approx \frac{1}{\sqrt{2}}(\bm e_x+i\lambda_2\bm e_y-\theta_{k_2} \bm\nu)\,,
\end{align}
where $\theta_{p}$, $\theta_{q}$, and $\theta_{k_2}$ are the polar angles of the vectors $\bm p$, $\bm q$, and $\bm k_2$. Within our approximation it is convenient to introduce the vectors $\bm \theta_{p}=\bm p_{\perp}/p$ etc. We remind that the orts $\bm e_x$ and $\bm e_y$ are directed along $\bm k_{2\perp}$ and $\bm k_1 \times \bm k_2$, respectively.

Using Eq.(\ref{spinors}) we obtain
\begin{eqnarray}\label{ab}
&&a_{+-}=1-\frac{\theta_{pq}^2}{8}+\frac{i}{4}\bm\nu\cdot[\bm\theta_p\times\bm\theta_q]\,,\quad  a_{-+}=-1+\frac{\theta_{pq}^2}{8}+\frac{i}{4}\bm\nu\cdot[\bm\theta_p\times\bm\theta_q]\,,\nonumber\\
&&a_{++}=-\frac{1}{\sqrt{2}}\bm e_-\cdot\bm\theta_{pq}\,,\quad  a_{--}=-\frac{1}{\sqrt{2}}\bm e_+\cdot\bm\theta_{pq}\,, \nonumber\\
&&\bm b_{+-}=\left[1-\frac{1}{8}(\bm\theta_{p}+\bm\theta_{q})^2-\frac{i}{4}\bm\nu\cdot[\bm\theta_p\times\bm\theta_q]\right]\bm\nu +\frac{1}{2}(\bm\theta_{p}+\bm\theta_{q})
+\frac{i}{2}[\bm\nu\times\bm\theta_{pq}]\,,\nonumber\\
&&\bm b_{-+}=\left[1-\frac{1}{8}(\bm\theta_{p}+\bm\theta_{q})^2+\frac{i}{4}\bm\nu\cdot[\bm\theta_p\times\bm\theta_q]\right]\bm\nu +\frac{1}{2}(\bm\theta_{p}+\bm\theta_{q})
-\frac{i}{2}[\bm\nu\times\bm\theta_{pq}]\,,\nonumber\\
&&\bm b_{++}=\frac{1}{\sqrt{2}}(\bm e_-,\bm\theta_{p}+\bm\theta_{q})\bm\nu-\sqrt{2}\bm e_-\,,\quad \bm b_{--}=-\frac{1}{\sqrt{2}}(\bm e_+,\bm\theta_{p}+\bm\theta_{q})\bm\nu+\sqrt{2}\bm e_+\,,
\end{eqnarray}
where $\bm\theta_{pq}=\bm\theta_{p}-\bm\theta_{q}$.

The main contribution to the integrals in Eq. (\ref{MD1}) is given by $r_{1,2}\sim \omega_1/m^2$, and by the impact parameters $r_{1\,\perp}\sim r_{2\,\perp}\sim 1/\Delta$. If $\Delta\gg m^2/\omega_1$ then the angle between the vectors $-\bm r_2$ and $\bm k_1$ is small. The angle between the vectors $\bm r_1$ and $\bm k_1$ may be either small or close to $\pi$ and we will call $M^{(1,1)}$ and $M^{(1,2)}$ the corresponding contributions to $M^{(1)}=M^{(1,1)}+M^{(1,2)}$.

For small angle between the vectors $\bm r_1$ and $\bm k_1$ one can use the quasiclassical form of the Green's function  $D(\bm r_1,\,\bm r_2|\varepsilon_{ p}+\omega_2)$ of the squared Dirac equation in the Coulomb field \cite{LMS2004}:
\begin{multline}\label{Dqc}
D(\bm r_1,\,\bm r_2|\varepsilon)=\frac{i\kappa}{8\pi^2r_1r_2}e^{i\kappa (r_1+r_2)}\!\int\! d\bm s\,\exp\left\{i\kappa \left[\frac{(r_1+r_2)}{2r_1r_2}s^2+\bm s\cdot\bm\theta_{12}\right]\right\}
\\
\times\left(\frac{s^2}{4r_1r_2}\right)^{-i\eta}
\left[1-\frac{1}{2}\bm\alpha\cdot\left(\frac{r_1+r_2}{r_1r_2}\bm s+\bm\theta_{12}\right)\right],
\end{multline}
where $\kappa=\sqrt{\varepsilon^2-m^2}$, $\bm s$ is the two-dimensional vector in the plane perpendicular to $\bm r_1-\bm r_2$ and  $\bm\theta_{12}=\bm r_1/r_1+\bm r_2/r_2$. We can also use the quasiclassical form of the wave function $v _{\bm q}^{(+)}(\bm r_2 )$ and the eikonal form of the wave function  $\bar u_{\bm p }^{(-)}(\bm r_1 )$:
\begin{align}\label{wfqc}
v _{\bm q}^{(+)}(\bm r_2 )&=\frac{q}{2i\pi r_2}e^{iqr_2}\!\int\! d\bm \tau\,\exp\left[iq\left(\frac{\tau^2}{2r_2}+\bm \tau\cdot\bm\theta_{2q}\right)\right]
\left(\frac{q\tau^2}{4r_2}\right)^{i\eta}
\left[1+\frac{1}{2}\bm\alpha\cdot\left(\frac{\bm \tau}{r_2}+\bm\theta_{2q}\right)\right]v_{\bm q}, \nonumber\\
\bar{u}_{\bm p}^{(-)}({\bm r}_1)&=\bar{u}_{\bm p}e^{-i\bm p\cdot\bm r_1}(pr_1)^{-i\eta}.
\end{align}
Here $\bm\theta_{2q}=-\bm r_{2}/r_2-\bm q/q$,  $\bm \tau$ is the two-dimensional vector in the plane  perpendicular to $\bm q$.
For small angle between the vectors $-\bm r_1$ and $\bm k_1$ one can use the eikonal form of the Green's function  $D(\bm r_1,\,\bm r_2|\varepsilon_{ p}+\omega_2)$ and the quasiclassical form of the  wave function  $\bar u_{\bm p }^{(-)}(\bm r_1 )$ \cite{LMS2004}:
\begin{align}\label{Deikuqc}
D(\bm r_1,\,\bm r_2|\varepsilon)&=-\frac{1}{4\pi r_{12}}e^{i\kappa r_{12}}\left(\frac{r_2}{r_1}\right)^{i\eta}\,,\quad r_2>r_1\,,\quad r_{12}=|\bm r_1-\bm r_2|\,,\nonumber\\
\bar{u}_{\bm p}^{(-)}(\bm r_1 )&=\frac{p}{2i\pi r_1}e^{ipr_1}\bar{u}_{\bm p}\!\int\! d\bm s\,\exp\left[ip\left(\frac{s^2}{2r_1}+\bm s\cdot\bm\theta_{1p}\right)\right]
\left(\frac{ps^2}{4r_1}\right)^{-i\eta}
\left[1-\frac{1}{2}\bm\alpha\cdot\left(\frac{\bm s}{r_1}+\bm\theta_{1p}\right)\right]\,,
\end{align}
where  $\bm s$ is the two-dimensional vector in the plane perpendicular to $\bm p$, and $\bm\theta_{1p}=-\bm r_1/r_1-\bm p/p$. The quasiclassical wave functions in (\ref{wfqc}) and (\ref{Deikuqc}) are the integral representations of the Furry-Sommerfeld-Maue wave functions \cite{Fu, ZM} (see also \cite{BLP82}). The most simple way to derive this integral representations is to use the relation between the wave functions and the Green function $D(\bm r_1,\,\bm r_2|\varepsilon)$, see \cite{DLMR2014}.

\section{Calculation of the matrix element}
To calculate the matrix element (\ref{MD1}) at $\Delta\gg \Delta_{\mathrm{min}}=m^2(\varepsilon_p+\varepsilon_q)/2\varepsilon_p\varepsilon_q$ we substitute the wave functions and the Green's function to Eq. (\ref{MD1}), take the trace, perform the expansion of the integrand in the phase and in the pre-exponent with respect to  small angles, taking into account  the leading terms, and then take the Gaussian  integrals  over $\bm\theta_{1p}$ and $\bm\theta_{2q}$. Note that, within our accuracy, $\bm s$, $\bm \tau$, $\bm\theta_{12}$, $\bm\theta_{1p}$, and $\bm\theta_{2q}$ are perpendicular to $\bm k_1$. Then we pass from the variables $\bm s$ and $\bm \tau$ (in the integral representation of the quasiclassical Green's function and the quasiclassical wave functions, or two quasiclassical wave functions) to the variables $\bm T=\bm\tau+\bm s$ and $\bm \xi=\bm\tau-\bm s$. After that both contributions $M^{(1,1)}_1$ and $M^{(1,2)}$ have the form
\begin{equation}\label{int}
M^{(1,i)}_{\lambda_1\lambda_2\mu_p\mu_q}=\int_0^\infty dr_2\int_0^{L_i} dr_1
\int d\bm T\int d\bm\xi \left(\frac{|\bm T+\bm\xi|}{|\bm T-\bm\xi|}\right)^{2i\eta} \exp\left[-\frac{i}{2}\bm T\cdot\bm\Delta_\perp\right]
{\cal G}_i(r_1,r_2,\bm\xi)\,,
\end{equation}
where ${\cal G}_{1,2}(r_1,r_2,\bm\xi)$ are some functions, $L_1=\infty$ and   $L_2=r_2$.
To perform further integration we use the transformation \cite{LMS1997}
\begin{eqnarray}\label{transform}
&&\int d\bm T  \left(\frac{|\bm T+\bm\xi|}{|\bm T-\bm\xi|}\right)^{2i\eta} \exp\left[-\frac{i}{2}\bm T\cdot\bm\Delta_\perp\right]=
\int d\bm T  \left(\frac{|\bm T+\bm\Delta_\perp|}{|\bm T-\bm\Delta_\perp|}\right)^{2i\eta} \exp\left[-\frac{i}{2}\bm T\cdot\bm\xi\right]\frac{\xi^2}{\Delta_\perp^2}\nonumber\\
&&=-\frac{4}{\Delta_\perp^2}\int d\bm T  \left(\frac{|\bm T+\bm\Delta_\perp|}{|\bm T-\bm\Delta_\perp|}\right)^{2i\eta} \bm \nabla_{\bm T}^2\exp\left[-\frac{i}{2}\bm T\cdot\bm\xi\right]\nonumber\\
&&=\frac{8i\eta}{\Delta_\perp^2}\int d\bm T  \left(\frac{|\bm T+\bm\Delta_\perp|}{|\bm T-\bm\Delta_\perp|}\right)^{2i\eta} \bm\chi\cdot \bm \nabla_{\bm T}\exp\left[-\frac{i}{2}
\bm T\cdot\bm\xi\right]\,,\nonumber\\
&&\bm\chi=\frac{\bm T+\bm\Delta_\perp}{(\bm T+\bm\Delta_\perp)^2}-
\frac{\bm T-\bm\Delta_\perp}{(\bm T-\bm\Delta_\perp)^2}\,.
\end{eqnarray}
After this transformation the integrals over $\bm\xi$, $r_1$, and $r_2$ can be easily taken, and we obtain for the total amplitude $M_{\lambda_1\lambda_2\mu_p\mu_q}=M_{\lambda_1\lambda_2\mu_p\mu_q}^{(1)}+
M_{\lambda_1\lambda_2\mu_p\mu_q}^{(2)}$:
\begin{align}\label{MD11}
M_{\lambda_1\lambda_2\mu_p\mu_q}&=\frac{32\eta}{\omega_1\omega_{2}\Delta^2}\int d\bm T  \left(\frac{|\bm T+\bm\Delta_\perp|}{|\bm T-\bm\Delta_\perp|}\right)^{2i\eta}\bm\chi \cdot \bm \nabla_{\bm T}\mathcal{F}_{\lambda_1\lambda_2\mu_p\mu_q}(\bm T)\,,\nonumber\\
\mathcal{F}_{\lambda_1\lambda_2\mu_p\mu_q}(\bm T)&={ F}_{\lambda_1\lambda_2\mu_p\mu_q}(\bm p,\bm q, \bm T)-
{ F}_{\lambda_1\lambda_2\mu_q\mu_p}(\bm q,\bm p, -\bm T)\,,
\end{align}
where $\bm\chi$ is defined in (\ref{transform}) and the functions ${F}_{\lambda_1\lambda_2\mu_p\mu_q}(\bm p,\bm q, \bm T)$ are
\begin{align}\label{F}
F_{+++-}&=-\left(\varepsilon_{p}+\omega_{2}\right)^{2}\bm{e}_+\cdot\left(\bm{T}-\bm{\delta}_{q}\right)\left(\bm{e}_-\cdot\bm{A}\right)\,,\nonumber\\
F_{+-+-}&=-\varepsilon_{p}\left(\varepsilon_{p}+\omega_{2}\right)\bm{e}_+\cdot\left(\bm{T}-\bm{\delta}_{q}\right)\left(\bm{e}_+\cdot\bm{A}\right)\,,\nonumber\\
F_{++-+}&=\varepsilon_{p}\varepsilon_{q}\bm{e}_+\cdot\left(\bm{T}-\bm{\delta}_{q}\right)\left(\bm{e}_-\cdot\bm{A}\right)
+2m^{2}\omega_1\omega_2B
-\frac{\varepsilon_{q}\varepsilon_{p}\omega_1\omega_2}{2\left(\varepsilon_{p}+\omega_{2}\right)D_{2}}\,,\nonumber\\
F_{+--+}&=\varepsilon_{q}\left(\varepsilon_{p}+\omega_{2}\right)\bm{e}_+\cdot\left(\bm{T}-\bm{\delta}_{q}\right)\left(\bm{e}_+\cdot\bm{A}\right)\,,\nonumber\\
F_{++++}&=\sqrt{2}m\left(\varepsilon_{p}+\omega_{2}\right)\omega_1\left(\bm{e}_-\cdot\bm{A}\right)\,,\nonumber\\
F_{++--}&=-\sqrt{2}m\left(\varepsilon_{p}+\omega_{2}\right)\omega_2\bm{e}_+\cdot\left(\bm{T}-\bm{\delta}_{q}\right)B\,,\nonumber\\
F_{+-++}&=\sqrt{2}m\varepsilon_{p}\omega_1\left(\bm{e}_+\cdot\bm{A}\right)
-\sqrt{2}m\varepsilon_{q}\omega_{2}\bm{e}_+\cdot\left(\bm{T}-\bm{\delta}_{q}\right)B\,,\nonumber\\
F_{+---}&=0\,,\nonumber\\
F_{\lambda_1\lambda_{2}\mu_{p}\mu_{q}}&=-\mu_{p}\mu_{q}\left(F_{\overline{\lambda}_{1}\overline{\lambda}_{2}\overline{\mu}_{p}\overline{\mu}_{q}}\right)^*\,.
\end{align}
Here $\overline{\mu}_{p,q}=-{\mu}_{p,q}$, $\overline{\lambda}_{1,2}=-{\lambda}_{1,2}$, and
\begin{gather}
\bm{A}=\frac1{D_1} \left[\frac{\varepsilon_{p}\bm{\theta}_{pk_2}}{2(m^{2}+\varepsilon_{p}^2\bm{\theta}_{pk_2}^2)}+
\frac{\omega_{2}\varepsilon_{q}\left(\bm{T}+\bm{\Delta_\perp}-2\varepsilon_{p}\bm{\theta}_{pk_2}\right)}{\left(\varepsilon_{p}+\omega_{2}\right)D_2}\right] \,,\\
B= \frac1{D_1} \left[\frac{1}{4(m^{2}+\varepsilon_{p}^2\bm{\theta}_{pk_2}^2)}-\frac{\omega_{2}\varepsilon_{q}}{\left(\varepsilon_{p}+\omega_{2}\right)D_2}\right]\,,\nonumber\\
D_1=4m^{2}+\left(\bm{T}-\bm{\delta}_{q}\right){}^{2}\,,\quad \bm{\delta}_{q}=\bm{\Delta_\perp}-2\varepsilon_{q}\bm{\theta}_{q}\,,
\quad
\bm \theta_{pk_2}=\bm\theta_{p}-\bm\theta_{k_2}\,,
\nonumber\\
D_2=\frac{4\omega_{1}\omega_{2}\varepsilon_{p}\varepsilon_{q}}{\varepsilon_{p}+\varepsilon_{q}}\bm{\theta}_{k_2}^{2}+\left(\varepsilon_{p}+\varepsilon_{q}\right)\left[\left(\bm{T}-\varepsilon_{p}\bm{\theta}_{p}+\varepsilon_{q}\bm{\theta}_{q}-\frac{\varepsilon_{p}-\varepsilon_{q}}{\varepsilon_{p}+\varepsilon_{q}}\omega_{2}\bm{\theta}_{k_2}\right)^{2}+4m^{2}\right]\,.
\end{gather}
In Eq. (\ref{MD11}) we have omitted for convenience  the inessential factor $(q/p)^{i\eta}$, and replaced $\Delta_\perp^2$ by $\Delta^2$ in the coefficient of Eq. \eqref{MD11}. After such replacement Eq. (\ref{MD11}) can be used not only at $\Delta_\perp\gg \Delta_z\sim m^2/\omega_1$  but  at $\Delta_\perp\sim \Delta_z$ as well, cf. \cite{LMS1997}. We remind that $d\bm T=dT_x\,dT_y$.

In Ref. \cite{CW1970} the impact-factor approach has been suggested. Using this approach, we have derived the amplitudes of the process under discussion and obtained the result, which is in agreement with our result \eqref{MD11}.

For  $\omega_2\ll p,\,q$ the expression (\ref{MD11}) is essentially simplified,
\begin{multline}\label{MDsp}
M_{\lambda_1\lambda_2\mu_p\mu_q}=\frac{16\eta}{\omega_1\omega_2\Delta^2}\left[\frac{\varepsilon_p^2(\bm e^*_{\lambda_2}\cdot\bm\theta_p)}{m^2+\varepsilon_p^2\theta_p^2}-
\frac{\varepsilon_q^2(\bm e^*_{\lambda_2}\cdot\bm\theta_q)}{m^2+\varepsilon_q^2\theta_q^2}\right]\\
\int d\bm T  \left(\frac{|\bm T+\bm\Delta_{\perp}|}{|\bm T-\bm\Delta_{\perp}|}\right)^{2i\eta}\bm\chi \cdot \bm \nabla_{\bm T}\frac{1}{4m^2+(\bm \delta_0-\bm T)^2}\\
\times \left[ \delta_{\mu_p,-\mu_q}(\varepsilon_p\delta_{\mu_p,\lambda_1}-\varepsilon_q\delta_{\mu_q,\lambda_1})(\bm e_{\lambda_1}, \bm \delta_0-\bm T)
+\sqrt{2}m\omega_1\lambda_1\delta_{\mu_q,\lambda_1}\delta_{\mu_p,\lambda_1}\right]\,,
\end{multline}
where $ \bm \delta_0=\varepsilon_p\bm\theta_{p}-\varepsilon_q\bm\theta_{q}$.
This result can also be obtained directly within the soft-photon-emission approximation \cite{BLP82}.

\section{Born amplitudes and Coulomb corrections}
Let us represent the amplitude $M$ as
\begin{equation}
M=M^{B}+M^{C}\,,
\end{equation}
where $M^{B}$ is linear in $\eta$ term (Born amplitude) and $M^{C}$ is the contribution of the higher-order terms (Coulomb corrections).
In order to find the Born term we  omit the factor $(|\bm T+\bm\Delta_\perp|/|\bm T-\bm\Delta_\perp|)^{2i\eta}$ and perform the integration by parts using the relation
\begin{equation}
 \bm \nabla_{\bm T} \cdot\bm\chi=2\pi[\delta(\bm T+\bm\Delta_\perp)-\delta(\bm T-\bm\Delta_\perp)]\,.
\end{equation}
As a result we obtain
\begin{equation}\label{MDB}
M^{B}_{\lambda_1\lambda_2\mu_p\mu_q}=\frac{64\pi \eta}{\omega_1\omega_2\Delta^2}[ \mathcal{F}_{\lambda_1\lambda_2\mu_p\mu_q}(\bm \Delta_\perp)-\mathcal{F}_{\lambda_1\lambda_2\mu_p\mu_q}(-\bm \Delta_\perp)].
\end{equation}
In order to derive the explicit expression for the Coulomb corrections we write
\begin{multline}
\bm\chi\cdot\bm\nabla_{\bm T} \mathcal{F}(\bm T)=
\frac{(\bm T+\bm\Delta_\perp)}{(\bm T+\bm\Delta_\perp)^2}\cdot \nabla_{\bm T} [\mathcal{F}(\bm T)-\mathcal{F}(-\bm \Delta_\perp)]\\ -
\frac{(\bm T-\bm\Delta_\perp)}{(\bm T-\bm\Delta_\perp)^2}\cdot \nabla_{\bm T} [\mathcal{F}(\bm T)-\mathcal{F}(\bm \Delta_\perp)]
\end{multline}
and perform integration by parts over $\bm T$ in Eq. \eqref{MD11}. The surface term gives the Born amplitude \eqref{MDB}, and the Coulomb corrections read
\begin{multline}\label{MDC}
M_{\lambda_1\lambda_2\mu_p\mu_q}^C=-\frac{128i\eta^2}{\omega_1\omega_{2}\Delta^2}\int \frac{d\bm T}{(\bm T+\bm\Delta_\perp)^2(\bm T-\bm\Delta_\perp)^2}  \left(\frac{|\bm T+\bm\Delta_\perp|}{|\bm T-\bm\Delta_\perp|}\right)^{2i\eta}\\
\times \left\{
(\bm\Delta_\perp^2+\bm T\cdot\bm{\Delta_\perp})[
\mathcal{ F}_{\lambda_1\lambda_2\mu_p\mu_q}(\bm T)-
\mathcal{ F}_{\lambda_1\lambda_2\mu_p\mu_q}(\bm \Delta_\perp)
]\right.\\
\left.
+
(\bm\Delta_\perp^2-\bm T\cdot\bm{\Delta_\perp})[
\mathcal{ F}_{\lambda_1\lambda_2\mu_p\mu_q}(\bm T)-
\mathcal{ F}_{\lambda_1\lambda_2\mu_p\mu_q}(-\bm \Delta_\perp)
]
\right\}\,.
\end{multline}
Note that it is possible to reduce the expression \eqref{MDC} to one-fold integral using the trick from Ref. \cite{LMS1998}.
However, the resulting formulas are very cumbersome and we do not present them here.

\section{Results and discussion}

Let us discuss the effect of screening. This effect is important only for small $\Delta\lesssim r^{-1}_{scr}\ll m$, where  $r_{scr}\sim m\alpha Z^{1/3}$ is the screening radius. For such  small
$\Delta$ the amplitude (\ref{MD11}) coincides with the Born amplitude at small $\Delta$, where the effect of screening  may be accounted for by multiplying the amplitude  $M^{B}_{\lambda_1\lambda_2\mu_p\mu_q}$ by an atomic form factor $[1-F_e(\Delta^2)]$. This form factor vanishes  at $\Delta=0$ and tends  to unity at $\Delta\rightarrow\infty$. A simple parametrization of this form factor can be found in  Ref. \cite{Mol47}. Thus, if we  multiply the amplitude  (\ref{MD11}) for the case of a pure Coulomb field by the atomic form factor $[1-F_e(\Delta^2)]$, we obtain the result which is valid in the atomic field for any $\Delta$.

\begin{figure}
\centering
\includegraphics[width=0.7\linewidth]{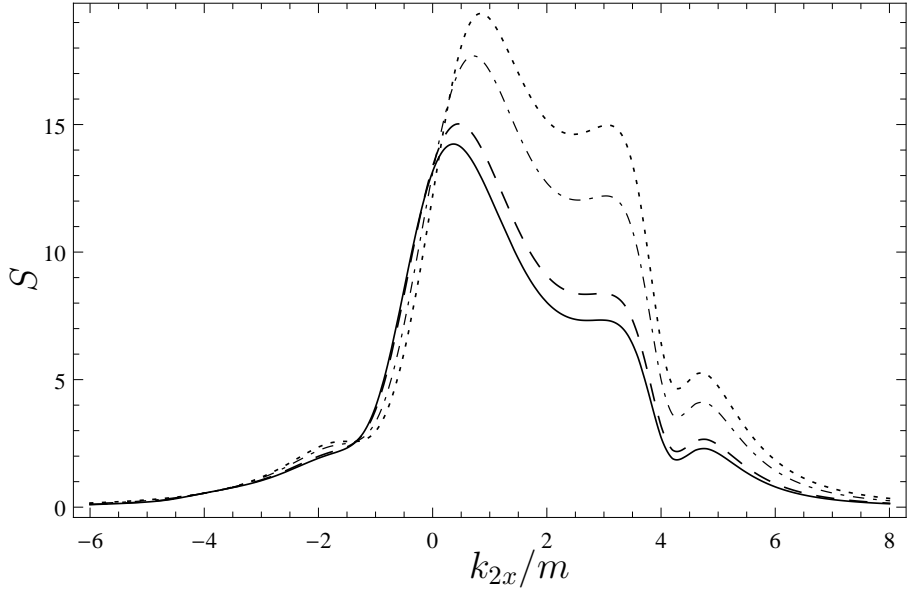}
\setlength{\unitlength}{0.7\linewidth}
\begin{picture}(0,0)
\put(-1.06,0.3){\large \rotatebox[]{90}{$S$}}
\put(-0.53,-0.03){\large $k_{2x}/m$}
\end{picture}
\caption{The quantity $S$, see Eq. \eqref{S}, as a function of $k_{2x}$ for $\varepsilon_p=0.4\omega_1$, $\varepsilon_q=0.25\omega_1$, $p_{x}=4.7 m$, $q_{x}=-0.8m$, $p_y=q_y=k_{2y}=0$; Born result (dotted curve), $Z=47$($\mathrm{Ag}$, dash-dotted curve), $Z=82$($\mathrm{Pb}$, dashed curve), and $Z=92$($\mathrm{U}$, solid curve).
}
\label{fig:cs1}
\end{figure}

\begin{figure}
\centering
\includegraphics[width=0.7\linewidth]{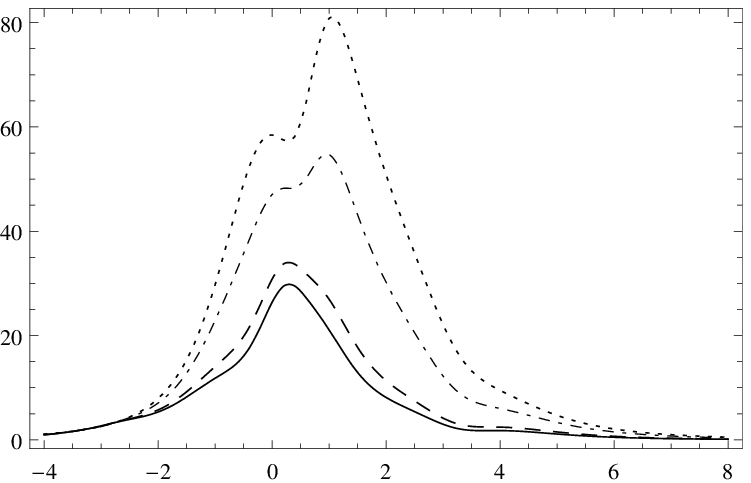}
\setlength{\unitlength}{0.7\linewidth}
\begin{picture}(0,0)
\put(-1.06,0.3){\large \rotatebox[]{90}{$S$}}
\put(-0.53,-0.03){\large $k_{2x}/m$}
\end{picture}
\caption{Same as Fig. \ref{fig:cs1} but for $p_{x}=0.7 m$, $q_{x}=2.33m$.
}
\label{fig:cs2}
\end{figure}

In order to demonstrate the importance of the Coulomb corrections in the process, we plot in Figs. \ref{fig:cs1} and  \ref{fig:cs2} the quantity $S$,
\begin{equation}\label{S}
S=\frac12\sum_{\lambda_1\lambda_2\mu_p\mu_q}\frac{\sigma_0^{-1}d\sigma_{\lambda_1\lambda_2\mu_p\mu_q}}{ d\bm{p}_\perp\,d\bm{q}_\perp\,d\bm{k}_{2\perp}d\varepsilon_pd\varepsilon_q}\,,\quad
\sigma_0=\frac{\alpha^2\eta^2\Delta_\perp^2}{(2\pi)^6m^6\omega_1\omega_2\Delta^4}
\end{equation}
as a function of $k_{2x}$ at fixed $\bm p_\perp$,  $\bm q_\perp$, $\varepsilon_p$,  $\varepsilon_q$, $k_{2y}=0$ and different values of the atomic charge number $Z$. For numerical calculations we used the two-fold integral representation \eqref{MDC}. In the vicinity of the point $\Delta_\perp=0$ ($k_{2x}=-3.9m$ in Fig. \ref{fig:cs1} and $k_{2x}=-3.03m$ in Fig. \ref{fig:cs2}), the Born result dominates over the Coulomb corrections as should be. However, it is seen that in general the Coulomb corrections significantly modify the cross section.

\begin{figure}
\centering
\includegraphics[width=0.7\linewidth]{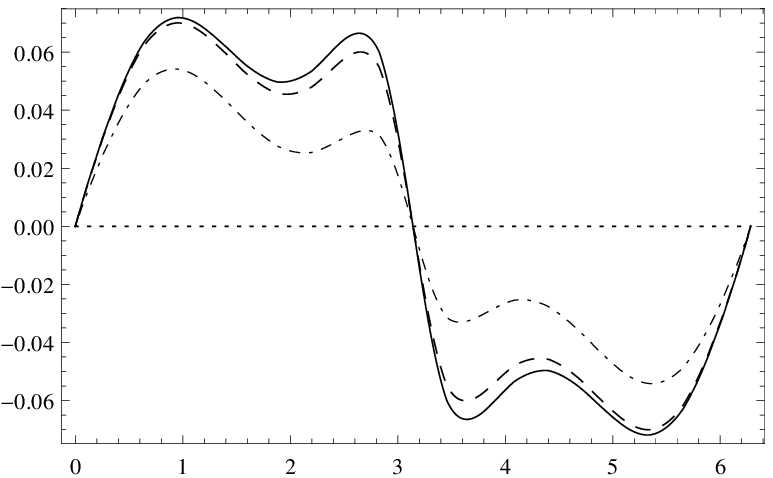}
\setlength{\unitlength}{0.7\linewidth}
\begin{picture}(0,0)
\put(-1.06,0.3){\large \rotatebox[]{90}{$\mathcal{A}$}}
\put(-0.53,-0.03){\large $\varphi$}
\end{picture}
\caption{Asymmetry $\mathcal{A}$, Eq. \eqref{as1}, as a function of angle $\varphi$ between $\bm k_{2\perp}$ and
$\bm p_\perp$ for
$\varepsilon_p=0.4\omega_1$, $\varepsilon_q=0.25\omega_1$, $\bm p_\perp\parallel-\bm q_\perp$, $p_{\perp}=4.7 m$, $q_{\perp}=0.8m$, $k_{2\perp}=m$;  Born result (dotted curve), $Z=47$($\mathrm{Ag}$, dash-dotted curve), $Z=82$($\mathrm{Pb}$, dashed curve), and $Z=92$($\mathrm{U}$, solid curve).
}
\label{fig:as1}
\end{figure}

\begin{figure}
\centering
\includegraphics[width=0.7\linewidth]{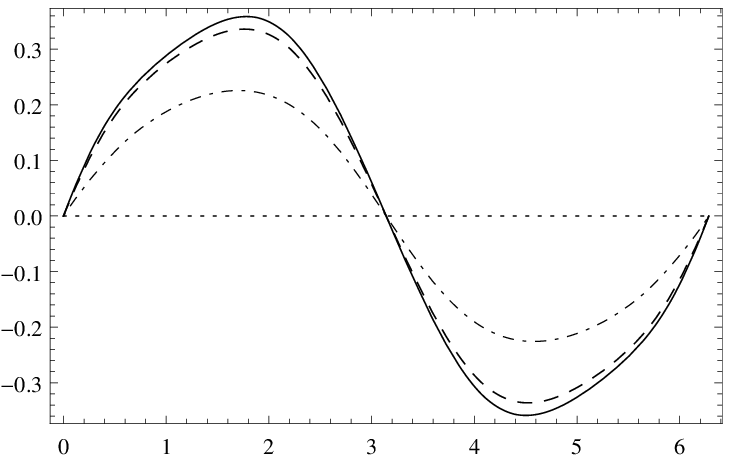}
\setlength{\unitlength}{0.7\linewidth}
\begin{picture}(0,0)
\put(-1.06,0.3){\large \rotatebox[]{90}{$\mathcal{A}$}}
\put(-0.53,-0.03){\large $\varphi$}
\end{picture}
\caption{Same as Fig. \ref{fig:as1} but for
$\bm p_\perp\parallel\bm q_\perp$, $p_{\perp}=0.7 m$, $q_{\perp}=2.33m$.
}
\label{fig:as2}
\end{figure}

There is an interesting question on the asymmetry $\mathcal{A}$ in the differential cross section for circularly polarized initial photon,
\begin{align}\label{as1}
\mathcal{A}&=\frac{d\sigma_+-d\sigma_-}{d\sigma_++d\sigma_-}\,,\nonumber\\
d\sigma_{\pm}&=\sum_{\lambda_2\mu_p\mu_q}d\sigma_{\pm\lambda_2\mu_p\mu_q}\,.
\end{align}
In the Born approximation the asymmetry vanishes for any $\bm p$, $\bm q$, and $\bm k_2$. This fact follows from the relation
\begin{equation}
M^B_{\lambda_1\lambda_{2}\mu_{p}\mu_{q}}=-\mu_{p}\mu_{q}\left(M^B_{\overline{\lambda}_{1}\overline{\lambda}_{2}\overline{\mu}_{p}\overline{\mu}_{q}}\right)^*\,,
\end{equation}
see Eqs. \eqref{F} and \eqref{MDB}.
However, for the Coulomb corrections this relation is not valid due to the complex factor $\left(\frac{|\bm T+\bm\Delta_\perp|}{|\bm T-\bm\Delta_\perp|}\right)^{2i\eta}$ in the integrand in Eq. \eqref{MDC}.
In Figs.  \ref{fig:as1} and  \ref{fig:as2} the asymmetry is shown as a function of the angle $\varphi$ between the vectors $\bm k_{2\perp}$ and  $\bm p_{\perp}$. As it should be, the asymmetry vanishes when $\bm k_1$, $\bm k_2$, $\bm p$, and $\bm q$ lie in the same plane ($\varphi=0,\pi$ in Figs. \ref{fig:as1} and  \ref{fig:as2}). It is seen that the asymmetry can reach tens of percent even for moderate values of $Z$.

\section{Conclusion}
Using the quasiclassical approximation, we have derived exactly in the parameter $\eta=Z\alpha$ the helicity amplitudes of $e^+e^-$ photoproduction  in the atomic field accompanied by bremsstrahlung. The results obtained, Eqs. \eqref{MD11}, \eqref{MDB}, \eqref{MDC},  have a compact form and are convenient for numerical calculations. They correspond to the leading high-energy small-angle asymptotics of the amplitude and have the relative uncertainty $\sim \max(\theta_p,\theta_q,\theta_{k_2},m/\omega_1)$. It is shown that, in general, the Coulomb corrections essentially modify the differential cross section as compared to the Born result. Moreover,when the initial photon is circularly polarized the Coulomb corrections lead to the asymmetry in the distribution over the azimuth angles $\varphi_i$ of produced particles with respect to the replacement $\varphi_i\to -\varphi_i$, Eq. \eqref{as1}.

\section*{Acknowledgments}
The work  has been supported in part by the Ministry of Education and Science of the Russian Federation and the RFBR grant no. 14-02-00016.

\end{document}